\pdfoutput=1
\documentclass[journal]{IEEEtran}

\usepackage{amsmath}
\usepackage{graphicx}
\usepackage{amssymb}
\usepackage{cite}
\usepackage{color}

\newcommand{\gPT}{\gamma_\mathrm{PT}}
\newcommand{\sgn}{\mathrm{sgn}}
\newcommand{\ii}{\mathrm{i}}
\renewcommand{\Re}{\mathrm{Re}}

\newcommand{\cP}{\ensuremath{\mathcal{P}}}
\newcommand{\cT}{\ensuremath{\mathcal{T}}}

\begin{document}

 \title{Solitary waves of a \cP\cT-symmetric Nonlinear Dirac equation}

\author{Jes\'us Cuevas--Maraver, Panayotis G. Kevrekidis, Avadh Saxena, Fred Cooper, Avinash Khare, Andrew Comech and Carl M. Bender%
\thanks{J. Cuevas--Maraver is with the Nonlinear Physics Group at Departamento de F\'{\i}sica Aplicada I, Universidad de Sevilla. Escuela Polit{\'e}cnica Superior, C/ Virgen de \'Africa, 7, 41011-Sevilla, Spain and the Instituto de Matem\'aticas de la Universidad de Sevilla (IMUS). Edificio Celestino Mutis. Avda. Reina Mercedes s/n, 41012-Sevilla, Spain (email:jcuevas@us.es).}%
\thanks{P.G. Kevrekidis is with the Department of Mathematics and Statistics, University of Massachusetts, Amherst, MA 01003-4515, USA (email: kevrekid@math.umass.edu).}%
\thanks{F. Cooper is with the Santa Fe Institute, Santa Fe, NM 87501, USA.}%
\thanks{A. Khare is with the Department of Physics, Savitribai Phule Pune University, Pune 411007, India.}%
\thanks{A. Comech is with the Department of Mathematics, Texas A\&M University, College Station, TX 77843-3368 and the Institute for Information Transmission Problems, Moscow 127994, Russia.}%
\thanks{C.M. Bender is with the Department of Physics, Washington University, St. Louis, MO 63130, USA.}%
\thanks{P.G. Kevrekidis, A. Saxena and F. Cooper are with the Center for Nonlinear Studies and Theoretical Division, Los Alamos National Laboratory, Los Alamos, New Mexico 87545, USA.}
}

\IEEEspecialpapernotice{(Invited Paper)}

\maketitle

\begin{abstract}
In the present work, we consider a prototypical example of a
{\cP \cT}-symmetric Dirac model. We discuss the underlying linear
limit of the model and identify the threshold of the {\cP \cT}-phase
transition in an analytical form. We then focus on the examination
of the nonlinear model. We consider the continuation in the
{\cP \cT}-symmetric model of the solutions of the corresponding
Hamiltonian model and find that the solutions can be continued
robustly as stable ones all the
way up to the {\cP \cT}-transition threshold. In the latter, they
degenerate into linear waves. We also examine the dynamics of
the model. Given the stability of the waveforms in the {\cP \cT}-exact
phase we consider them as initial conditions for parameters
outside of that phase. We find that both oscillatory dynamics and
exponential growth may arise, depending on the size of the corresponding
``quench''. The former can be characterized by
an interesting form of bi-frequency solutions that have been
predicted on the basis of the SU$(1,1)$ symmetry. Finally,
we explore some special, analytically tractable, but
not {\cP \cT}-symmetric solutions in the
massless limit of the model.
\end{abstract}


\section{Introduction}

The study of open systems bearing gain and loss
(especially so in a balanced form) is a topic that has
emerged over the past two decades as a significant
theme of study~\cite{Bender_review,special-issues,review}.
While the realm of {\cP \cT}-symmetry introduced by
Bender and collaborators was originally intended as
an alternative to the standard Hermitian quantum
mechanics, its most canonical realizations (beyond
the considerable mathematical analysis of the theme
in its own right at the level of operators and spectral
theory in mathematical physics) emerged elsewhere
in physics. More specifically, in optical systems~\cite{Muga,PT_periodic}
the analogy of the paraxial approximation of Maxwell's equations
and of the Schr{\"o}dinger equation formed the basis on which
the possibility of {\cP \cT}-symmetric realizations initially
in optical waveguide experiments was proposed and then
experimentally implemented~\cite{experiment}. The success
of this program motivated further additional initiatives
in other directions of experimental interest, including,
but not limited to, {\cP \cT}-symmetric electronic
circuits~\cite{tsampikos_recent,tsampikos_review}, mechanical
systems~\cite{pt_mech} and whispering-gallery microcavities~\cite{pt_whisper}.

Another theme of research that has been receiving increasing attention
recently, both in  the physics  and
in the mathematics community is that of the nonlinear Dirac equations (NLDEs).
While such models were proposed in the context of
high-energy physics over 50 years ago~\cite{thirring,gavri}, they
have, arguably, been far less widespread than their non-relativistic
counterpart~\cite{nogami}, the nonlinear Schr{\"o}dinger (NLS)
equation~\cite{sulem,ablowitz}. In recent years, however,
there has been a surge of activity around NLDE models fueled
to some extent by analytical solutions and computational
issues arising in associated numerical
simulations~\cite{sihong_rev,cooper,niur_recent,MR2892774}, as well as by
the considerable progress achieved by rigorous techniques towards
aspects of the spectral, orbital and asymptotic stability of solitary
wave solutions of such models~\cite{boucuc,DEP1,DEP2,DEP3,arXiv:1407.0606,linear-a}
and towards criteria for their spectral stability~\cite{comgus,berko}.
 Although our emphasis herein will be on the so-called
Gross-Neveu model~\cite{gn3} (sometimes also referred to as
the Soler model~\cite{soler}), we also mention in passing
that another main stream of activity in this direction has been
towards the derivation of NLDEs in the context e.g. of
bosonic evolution~\cite{carr1,carr2}
(or light propagation~\cite{ablowitz3,ablowitz4}) in
honeycomb optical lattices. In the latter context,
contrary to what we will be focusing on below, there is no
nonlinear interaction between the fields (of the two-component
spinor).

Our aim in the present work is to connect these
two budding areas of research, namely to propose
a prototypical {\cP \cT}-symmetric nonlinear
Dirac equation model (PT-NLDE).
Motivated by the considerable volume of activity, as well
as analytical availability of solutions within the Hamiltonian
limit, we will focus on the Gross-Neveu (or Soler) model.
In the next section, we will present the mathematical
formulation of the model. We will analyze its linear
limit and discuss the existence of a {\cP \cT}-symmetry
breaking critical point, i.e., the point of a {\cP \cT}-phase
transition. Then, we will turn to the nonlinear
variant of the model exploring the conditions for
the existence of a standing wave solution, as well as
discussing the linear (spectral) stability setup for such
a solution. Finally, we will briefly touch upon the fate
of the conservation laws, such as the power (squared
$L^2$ norm) and the energy. As a remarkable feature,
we find that the energy remains {\it invariant} within
the nonlinear PT-NLDE model, a feature that certainly
distinguishes the model from its NLS counterpart.
In Section III we will examine the numerical properties
of the standing wave solutions. Remarkably, we will find that
these standing wave solutions are stable {\it throughout} their interval
of existence tending to a linear limit of vanishing
amplitude as the linear threshold (i.e., the threshold of
the underlying linear model) of the {\cP \cT}-transition
is approached. Since no instability is encountered within
the exact {\cP \cT}-phase, we consider the propagation
beyond the relevant critical point (i.e., under
a quench in the gain-loss parameter $\gamma$) finding (a) the possibility
of oscillatory motion that we identify with a bi-frequency
state and connect to the invariances of the model and
(b) the possibility of exponential growth. Lastly, a
special case of vanishing mass solutions is analytically
identified and also numerically examined. The remarkable
feature in this case is that these solutions do {\it not}
respect the {\cP \cT}-symmetry.
In Section IV,
we summarize our findings and present some interesting
directions for future study.

\section{Model equations}

The system of choice in the present context will be the Gross-Neveu model in its generalized {\cP \cT}-symmetric (PT-NLDE) form:
\begin{IEEEeqnarray}{l}\label{eq:dyn}
\ii \partial_t U = \partial_x V - g (|U|^2-|V|^2)^k U + m U + \ii \gamma V, \IEEEyesnumber \IEEEyessubnumber \\
\ii \partial_t V = -\partial_x U + g (|U|^2-|V|^2)^k V - m V + \ii \gamma U. \IEEEyessubnumber
\end{IEEEeqnarray}
We will restrict our considerations to the one-dimensional setting,
as is evident from the above.
Eqs. (\ref{eq:dyn}) are {\cP \cT}-symmetric because they are invariant under the
transformation {\cP}: $x \rightarrow -x$, $U \rightarrow U$,
$V \rightarrow -V$ and {\cT}: $t \rightarrow -t$, $i \rightarrow -i$,
$U \rightarrow U$, and $V \rightarrow V$. This transformation assumes that
$U(x,t)$ is spatially even and that $V(x,t)$ is spatially odd.
Without loss of generality, we will set the mass $m=1$,
except when explicitly indicated otherwise,
while we will also use $g=1$ (the coefficient of the nonlinear term can
be scaled out, hence, when it is
present, we only care about its sign). The key
addition in this model in comparison to the earlier proposal
of~\cite{cooper} is the inclusion of the gain-loss term
proportional to $\gamma$ in the (implicit) form of the Dirac
matrix $\gamma_5$ (cf.~\cite{Bender_review}) multipliying the
spinor $(U,V)^T$.
In our case of two-component spinors,
the role of $\gamma_5$ is played by the Pauli matrix $\sigma_1$.
We note in passing that in its linear form,
the model can be converted under a suitable transformation
(associated with the so-called $\mathcal{C}$-operator)
to a Hamiltonian one with a reduced mass of
$\sqrt{m^2-\gamma^2}$~\cite{Bender_review,jones}. Although
the linear version of the model was proposed and analyzed in these works,
to the best of our knowledge, there has not been any previous
work in the realm of the nonlinear variant i.e., of the \cP\cT-NLDE.

It is straightforward to see that in the linear case
(of $g=0$), plane waves $U= A e^{i (\kappa x - \omega t)}$
and $V=i B e^{i (\kappa x - \omega t)}$ are solutions
provided the dispersion relation
$\omega= \pm \sqrt{m^2 + \kappa^2 - \gamma^2}$ is satisfied.
Not only does the above formula have the characteristic
Dirac form, but it also is consistent with the equivalence
of the linear {\cP \cT}-Dirac equation
with effective mass $\tilde{m}=\sqrt{m^2 - \gamma^2}$, as per the above discussion.

Having examined the linear states (plane waves) of the model,
let us now turn to the nonlinear ones, and more specifically
to  standing waves.
The relevant coherent structures will be of the form:

\begin{equation}\label{eq:stat}
    U(x,t)=\exp(-i \Lambda t) u(x), \qquad V(x,t)=\exp(-i \Lambda t) v(x)\,.
\end{equation}

Once such standing wave solutions are calculated (e.g., by fixed point methods
as will be discussed in the next section), their linear stability is considered by means of a Bogoliubov-de Gennes linearized stability analysis.
We note here, in passing, that unfortunately, contrary to what
is the case for the Hamiltonian limit of the model with $\gamma=0$,
for which explicit solutions exist as:
\begin{IEEEeqnarray}{rcl}\label{eq:solcont}
    u(x)&=&\sqrt{\frac{(m+\Lambda)\cosh^2(k\beta x)}{m+\Lambda\cosh(2k\beta x)}\left[\frac{(k+1)\beta^2}{g^2(m+\Lambda\cosh(2k\beta x))}\right]^{1/k}}, \IEEEyesnumber \IEEEyessubnumber\\
    v(x)&=&\sgn(x)\sqrt{\frac{(m-\Lambda)\sinh^2(k\beta x)}{m+\Lambda\cosh(2k\beta x)}\left[\frac{(k+1)\beta^2}{g^2(m+\Lambda\cosh(2k\beta x))}\right]^{1/k}} ,
    \IEEEyessubnumber
\end{IEEEeqnarray}
(where $\beta= \sqrt{m^2-\Lambda^2}$)
in the present PT-NLDE we have been unable to identify such
explicit solutions (with a notable exception for $m=0$ discussed
separately below). I.e, with the exception of the $m=0$ solutions given
below, we have not been able to identify additional analytical
solitary wave solutions of the case with $\gamma \neq 0$.
In the same vein, it does not appear to be
straightforward to generalize the transformation of~\cite{Bender_review,jones}
to the present nonlinear setting.

We now consider the linearization of the standing wave solutions
in order to extract information about their spectral stability.
More specifically, considering small perturbations [of order ${\rm O}(\delta)$, with $0< \delta \ll 1$] of the standing wave solutions, we substitute the ansatz

\begin{IEEEeqnarray}{l}
    U(x,t)=e^{-\ii \Lambda t} \left[u(x) + \delta (a_1(x) e^{\ii \omega t} + b_1(x)^{*} e^{-\ii \omega^{*} t}) \right], \IEEEyesnumber \IEEEyessubnumber \\
    V(x,t)=e^{-\ii \Lambda t} \left[v(x) + \delta (a_2(x) e^{\ii \omega t} + b_2(x)^{*} e^{-\ii \omega^{*} t}) \right] \IEEEyessubnumber
\end{IEEEeqnarray}
into Eqs.~(\ref{eq:dyn}), and then solve the ensuing [to O$(\delta)$] eigenvalue problem $\omega(a_1,a_2,b_1^*,b_2^*)^T=\mathcal{M}(a_1,a_2,b_1^*,b_2^*)^T$ with $\mathcal{M}$ being
\begin{equation}\label{eq:stabmat}
    \mathcal{M}=\left(\begin{array}{cc} L_1 & L_2 \\ \\ -L_2^* & -L_1^* \end{array}\right)-
    \ii\gamma\left(\begin{array}{cc} J & 0 \\ \\ 0 & J \end{array}\right)
\end{equation}
and
\begin{IEEEeqnarray}{lcl}
    L_1 &=&
    \left(\begin{array}{cc}
    g(|u|^2-|v|^2)^k-m+\Lambda & -\partial_x \\ \\
    \partial_x & -g(|u|^2-|v|^2)^k+m+\Lambda
    \end{array}\right)
    \IEEEnonumber \\
    &+&
    gk(|u|^2-|v|^2)^{k-1}
    \left(\begin{array}{cc}
    |u|^2 & -uv^* \\ \\
    -u^*v & |v|^2
    \end{array}\right), \IEEEnonumber \\ \\
    L_2 &=& gk(|u|^2-|v|^2)^{k-1}
    \left(\begin{array}{cc}
    u^2 & -uv \\ \\
    -uv & v^2
    \end{array}\right) , \IEEEnonumber \\ \\
    J &=& \left(\begin{array}{cc}
    0 &
    I \\ \\
    I &
    0
    \end{array}\right) , \IEEEnonumber
\end{IEEEeqnarray}
where $I$ is the identity matrix. The potential existence of an eigenvalue
with non-vanishing real part (i.e., an eigenfrequency $\omega$ with
non-vanishing imaginary part) suggests the existence of a
dynamical instability. If all the eigenvalues are imaginary (i.e.,
all eigenfrequencies $\omega$ are real), then the solution is spectrally
(neutrally) stable.

Finally, once the exact waveforms and their linear stability are identified, the corresponding full nonlinear dynamics of the
scheme is monitored by means of the solution of Eqs.~(\ref{eq:dyn})
in our numerical considerations of the next section.
A natural, theoretically motivated aspect to consider in that regard
is the (potential) preservation, by the numerical scheme, of the different
conservation laws. To that effect, we examine the fate of the
prototypical conservation laws (such as the power and the energy)
in the context of our PT-NLDE model.

Based on the power density,
\begin{equation}\label{eq:density}
    \rho(x)=|U(x,t)|^2+|V(x,t)|^2,
\end{equation}
we can define the charge (or power, as it is also referred
e.g. in the context of optics)
\begin{equation}\label{eq:charge}
    Q=\int\rho(x)\mathrm{d}x\,.
\end{equation}

From the dynamical equations (\ref{eq:dyn}) it is
straightforward to show that the charge is not preserved. Instead,
the following ``moment equation'' is satisfied:

\begin{equation}\label{eq:chargederiv}
    \frac{dQ}{dt}=4\gamma\int\Re(UV^*)\mathrm{d}x\,.
\end{equation}
Note that in
the case of a standing wave state, $dQ/dt=0$ and charge is conserved.

Although the charge is not generally conserved, remarkably
there is a conserved quantity in the form of the energy:

\begin{IEEEeqnarray}{rl}\label{eq:energy}
    E=\int  & \Big{[} \Re(U^*\partial_x V-V^*\partial_x U)+m(|U|^2-|V|^2) \IEEEnonumber \\
            &  -\frac{g}{k+1}[|U|^2-|V|^2]^{k+1}\Big{]}\mathrm{d}x\,.
\end{IEEEeqnarray}

Notice that there is no $\gamma$ dependence in this formula. That is, this is
the same definition for the energy as in the $\gamma=0$ limit. But,
intriguingly, $dE/dt=0$ even for $\gamma\neq0$. This is rather unusual
in our experience in {\cP \cT}-symmetric models and is effectively
related to the special form of introducing ${\cP \cT}$-symmetry
through the $\gamma_5$ matrix. We note that in this form, it is not
transparent (as it is e.g. in Schr{\"o}dinger {\cP \cT}-symmetric
models~\cite{Bender_review}) which component corresponds to the gain
and which one to the loss. Effectively, isolating the time-dependence
and the $\gamma$-dependent term in the equations (i.e.,
$\ii \partial_t U = \ii \gamma V$ and $\ii \partial_t V= \ii \gamma U$ and
momentarily ignoring the rest of the terms),
it appears as if both components bear both gain and loss. Nevertheless,
this $\gamma$-independent
effective energy conservation that we will numerically corroborate
below is certainly worth additional examination in order to
determine its origin.

\section{Numerical results}

\subsection{Massive NLD equation with $k=1$}

In the numerical computations presented herein,
we have utilized spectral collocation methods in order to approximate the spatial derivatives of Eq.~(\ref{eq:dyn}). As discussed in~\cite{ourprev},
the Chebyshev collocation is, arguably, the most suitable method
for approximating the relevant derivatives as it
gives a better spectral accuracy. However, because of the non-Hermiticity
of the system, the implementation of fixed-point methods, such as the
Newton method used herein, requires a high amount of computer memory
and more than $\sim500$ grid points which, in turn,
poses  implementation challenges. Furthermore, the method
has the drawback that the double humped solitons
(which occur for $\Lambda\leq1/3$ when $k=1$), cannot be well resolved
(i.e. the humps cannot be observed) because of the Chebyshev collocation
including more points at the edge of the system in comparison to the center.
Consequently, in what follows, a Fourier collocation scheme
has been implemented.
In suitable limit cases, we have checked that the results are similar
for the different implementations and that no extra spurious eigenmodes
arise in comparison to the Hermitian case. We note that hereafter
we will focus on the case of $k=1$ for our numerical
implementation.

\begin{figure}
\begin{tabular}{cc}
\includegraphics[width=4cm]{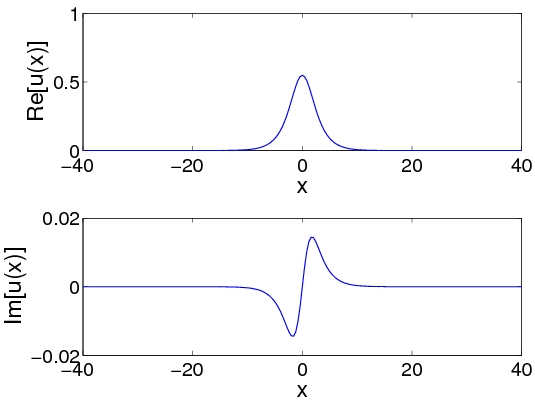} &
\includegraphics[width=4cm]{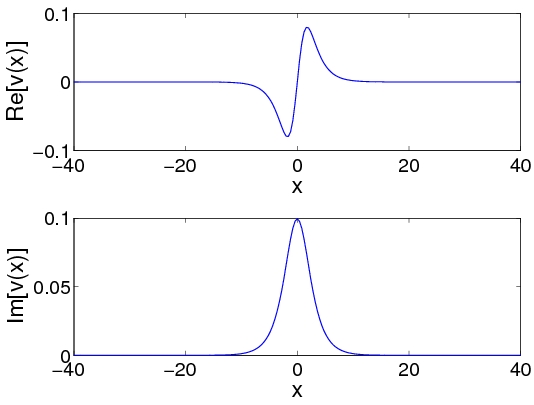} \\
\end{tabular}
\caption{Real and imaginary (top and bottom panels)
part of the spinor components (left and right panels)
for $\Lambda=0.8$ and $\gamma=0.3$.}
\label{fig:profile}
\end{figure}

\begin{figure}
\begin{tabular}{cc}
\includegraphics[width=4cm]{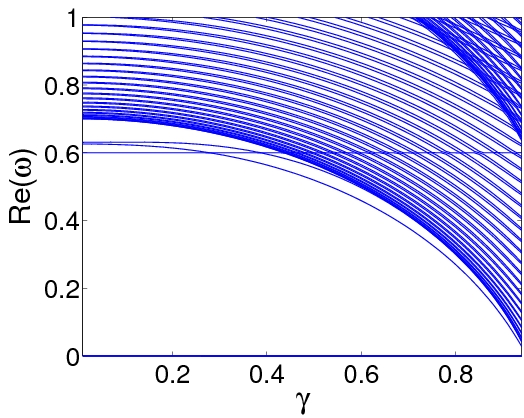} &
\includegraphics[width=4cm]{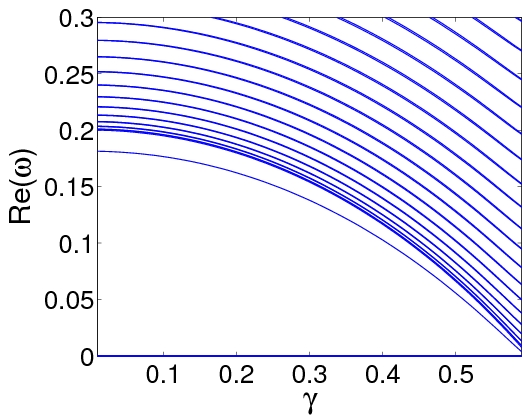} \\
\end{tabular}
\caption{Real part of the eigenfrequencies dependence for $\Lambda=0.3$ (left) and $\Lambda=0.8$ (right). Notice the existence on the left panel
of a constant frequency at $2\Lambda=0.6$. Notice also the approach of
the frequencies towards $0$, as per the discussed collision with the
linear limit (eigenfunctions) of the problem.}
\label{fig:stab}
\end{figure}

The first numerical result found by studying the standing wave solution is the
{\cP \cT} transition point. We have checked that this transition takes place when $\gamma=\gPT$ with $\gPT=\sqrt{m^2-\Lambda^2}$ (as we have taken $m=1$, $\gPT=\sqrt{1-\Lambda^2}$); notice that this is
consonant with our analytical prediction from the previous section
in the case of wavenumber $\kappa=0$.
Figure~\ref{fig:profile} shows the profile of a typical soliton with nonzero $\gamma$. Importantly,
we have observed (see Fig. \ref{fig:stab}) that the non-Hermiticity
does not introduce instabilities to our system.
The relevant spectrum features a zero eigenvalue of
algebraic multiplicity four and geometric multiplicity two.
This is present
in the spectrum of the linearized equation
due to the U(1) symmetry and due to
the translational symmetry, which are both preserved
when $\gamma\ne 0$, hence both the algebraic and geometric
multiplicity of this eigenvalue are preserved for all values of
$\gamma$, as is the presence of two generalized eigenvectors.
The spectrum also features the eigenfrequencies $\omega= \pm 2\Lambda$~\cite{MR2892774}
which can not leave the imaginary axis
since their presence in the spectrum
is due to
the SU(1,1) invariance (see the more detailed discussion below),
which is also preserved for any $\gamma$; the relevant eigenfrequency, which persists under variations of $\gamma$, can be discerned in the
left panel of Fig.~\ref{fig:stab}. For the rest of the spectrum
we note that according to \cite{linear-a},
 eigenfrequencies with nonzero imaginary part
can only be born
in the interval $\big(-1-|\Lambda|,1+|\Lambda|\big)$.
Here, however, all the eigenfrequencies remains inside this interval for all $\gamma$, as illustrated in Fig.~\ref{fig:stab},
solely tending towards $0$, as $\gamma$ approaches $\gPT$.

\begin{figure}
\begin{tabular}{cc}
\includegraphics[width=4cm]{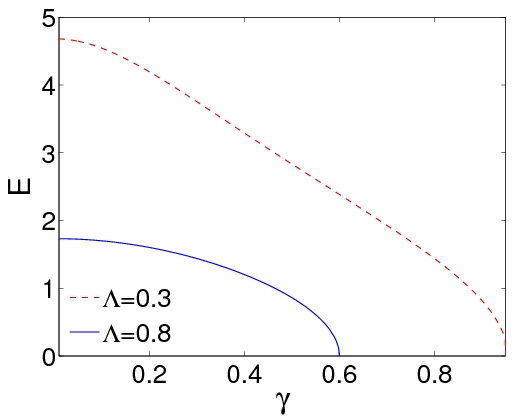} &
\includegraphics[width=4cm]{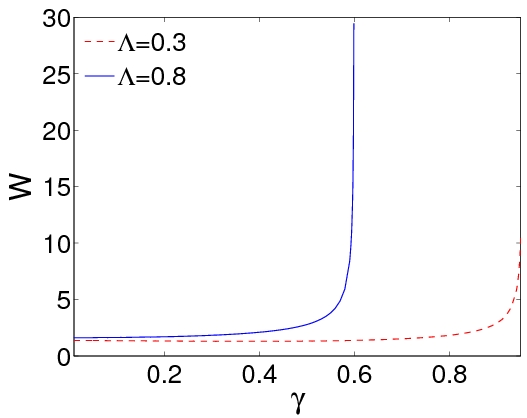} \\
\multicolumn{2}{c}{\includegraphics[width=4cm]{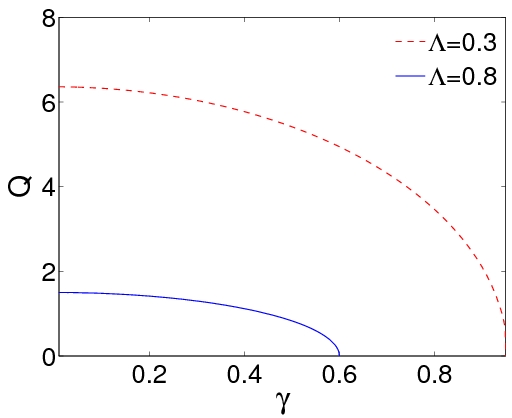}} \\
\includegraphics[width=4cm]{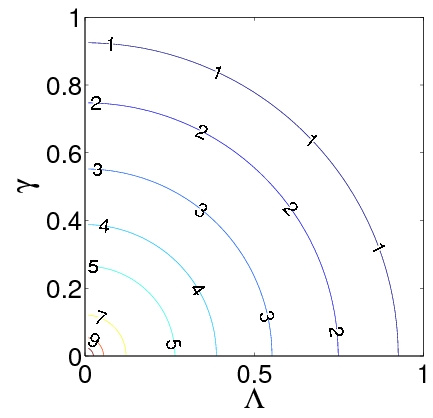} &
\includegraphics[width=4cm]{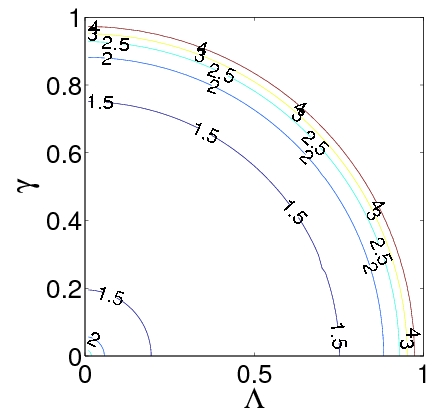} \\
\end{tabular}
\caption{The standing waves' energy $E$, width $W$ and charge
$Q$ dependence for $\Lambda=0.3$ and $\Lambda=0.8$ are shown in the first three panels. The bottom
panels present a two-parameter diagram of the dependence of the energy (bottom left) and width on the frequency $\Lambda$ and the gain-loss parameter $\gamma$.}
\label{fig:charge}
\end{figure}

Notice that the {\cP\cT}
transition is caused by the nonlinear solutions colliding with (or
degenerating into) linear modes. This fact can also be confirmed in the charge versus $\gamma$ plot of Fig. \ref{fig:charge}, where the charge (norm) and energy tend to zero (while the width of the solution diverges) when the transition
point is reached (actually, we have not been able to reach this point
exactly, as the soliton width increases drastically when approaching this
point).
It should be noted here that this is a distinct phenomenology
(again) in comparison to the NLS counterpart of the model. In the latter,
typically at the {\cP \cT}-phase transition a stable (center) and an
unstable (saddle) solution collide and disappear in a saddle-center
bifurcation. Here, a fundamentally different scenario arises
through the degeneration of the nonlinear modes into linear ones.
In the bottom panel of Fig.~\ref{fig:charge}, we provide two-parameter
diagrams of the relevant solutions as a function of the frequency
$\Lambda$ and the gain-loss parameter $\gamma$. The dependences
strongly suggest a ``combined'' monoparametric dependence on
$\gamma^2 + \Lambda^2$, although we have not been able to
analytically identify solutions bearing this dependence, except
in the limit of $m=0$; see below. For this reason (the strong similarity
of the dependence of monoparametric plots on $\Lambda$ and $\gamma$),
we do not show separately the dependence on $\Lambda$ or fixed
$\gamma$.

\begin{figure}
\begin{tabular}{cc}
{\footnotesize $\gamma_s=0.80$} & {\footnotesize $\gamma_s=0.90$} \\
\includegraphics[width=4cm]{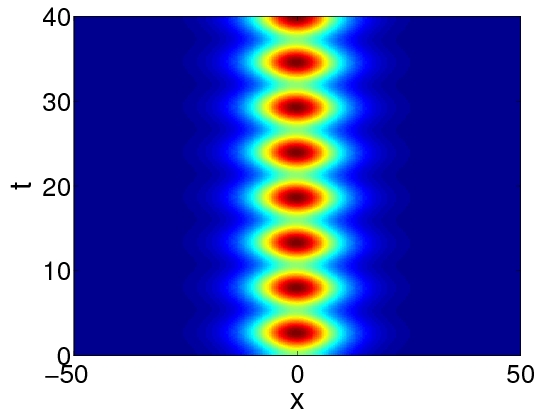} &
\includegraphics[width=4cm]{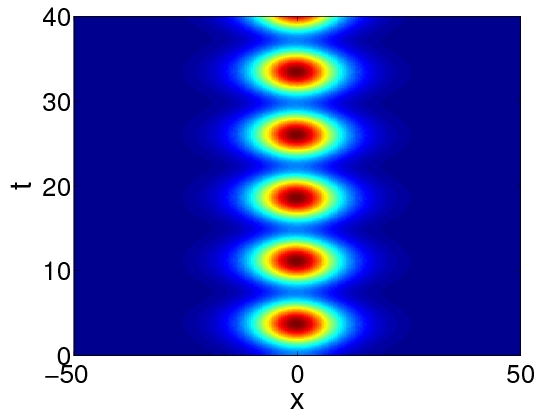} \\
{\footnotesize $\gamma_s=0.95$} & {\footnotesize $\gamma_s=0.99$} \\
\includegraphics[width=4cm]{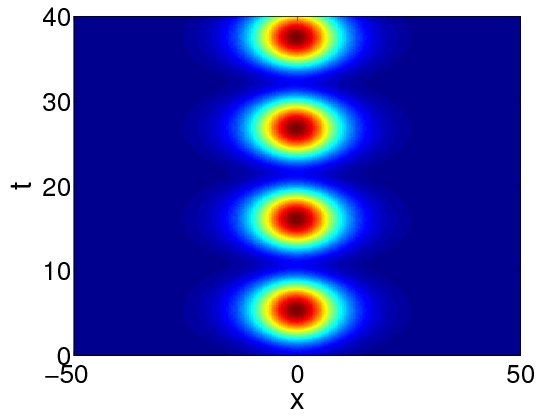} &
\includegraphics[width=4cm]{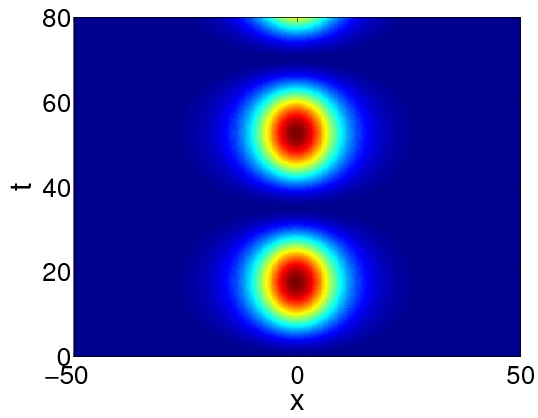} \\
\end{tabular}
\begin{tabular}{cc}
{\footnotesize $\gamma_s=0.80$} & {\footnotesize $\gamma_s=0.90$} \\
\includegraphics[width=4cm]{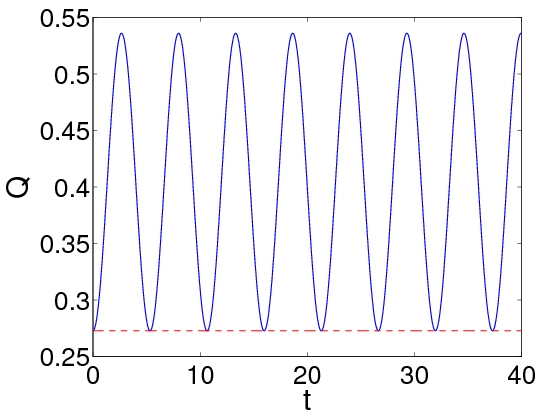} &
\includegraphics[width=4cm]{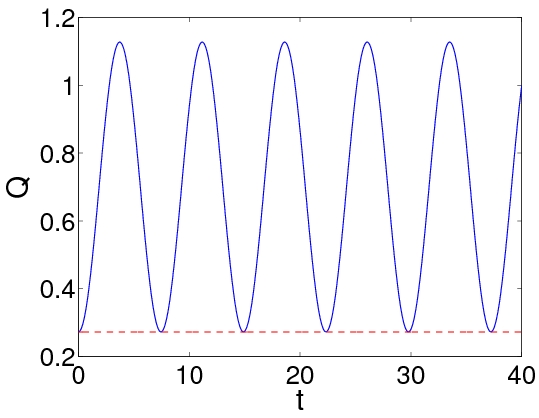} \\
{\footnotesize $\gamma_s=0.95$} & {\footnotesize $\gamma_s=0.99$} \\
\includegraphics[width=4cm]{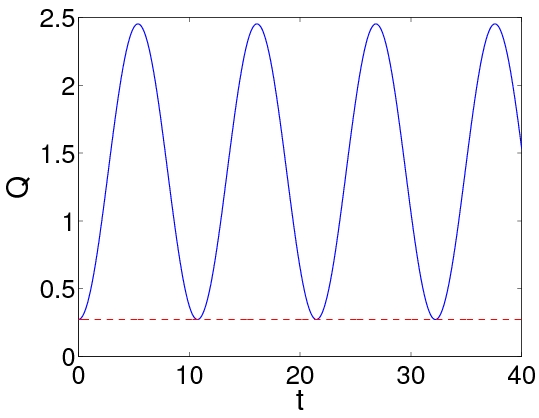} &
\includegraphics[width=4cm]{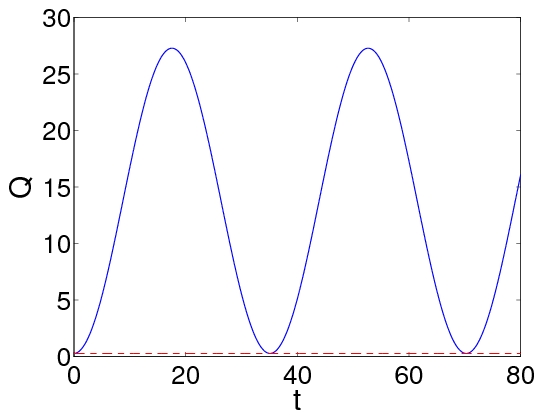} \\
\end{tabular}
\caption{Top two rows: Dynamics for $\Lambda=0.8$ using as initial condition the soliton with $\gamma=0.59$, but evolving Eq.~(\ref{eq:dyn}) for
$\gamma=\gamma_s$, shown in the respective panels.
Each panel shows the contour plot of the
space-time evolution of the soliton density.
Bottom two rows: same as the top but with  each panel showing
the time evolution of the soliton charge. The dashed
line corresponds to the charge of the initial condition.}
\label{fig:simul1}
\end{figure}

\begin{figure}
\begin{tabular}{cc}
{\footnotesize $\gamma_s=0.80$} & {\footnotesize $\gamma_s=0.90$} \\
\includegraphics[width=4cm]{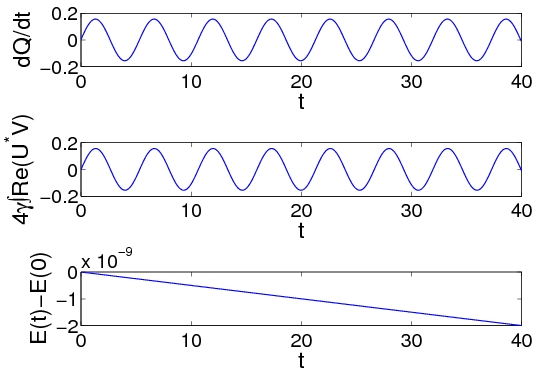} &
\includegraphics[width=4cm]{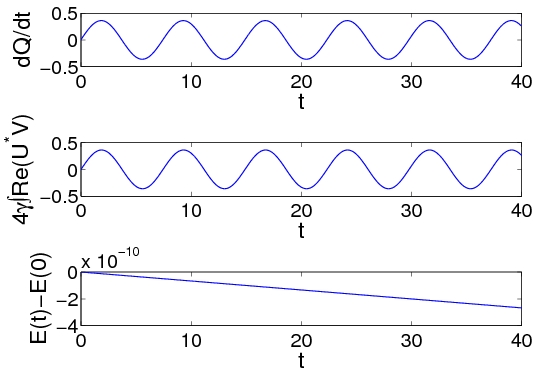} \\
{\footnotesize $\gamma_s=0.95$} & {\footnotesize $\gamma_s=0.99$} \\
\includegraphics[width=4cm]{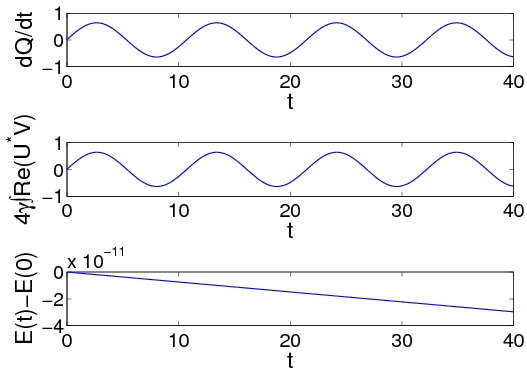} &
\includegraphics[width=4cm]{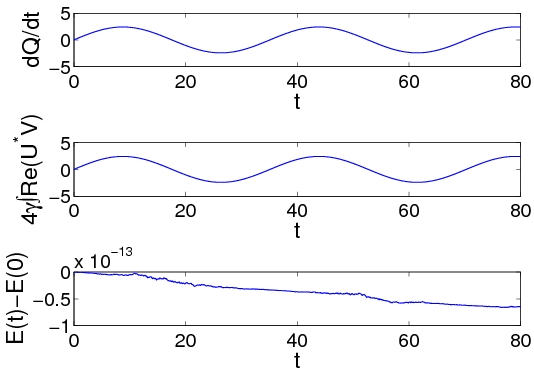} \\
\end{tabular}
\caption{Dynamics for $\Lambda=0.8$ using as initial condition the soliton with $\gamma_0=0.59$. Each panel shows the time evolution of the time derivative of the soliton charge and compares it with the right hand side of Eq. (\ref{eq:chargederiv}); the bottom panels for each example
of $\gamma$ illustrate the time dependence of the energy fluctuations.}
\label{fig:simul2b}
\end{figure}

\begin{figure}
\begin{tabular}{cc}
\includegraphics[width=4cm]{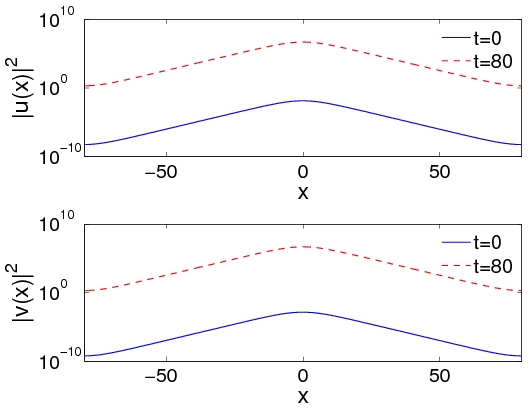} &
\includegraphics[width=4cm]{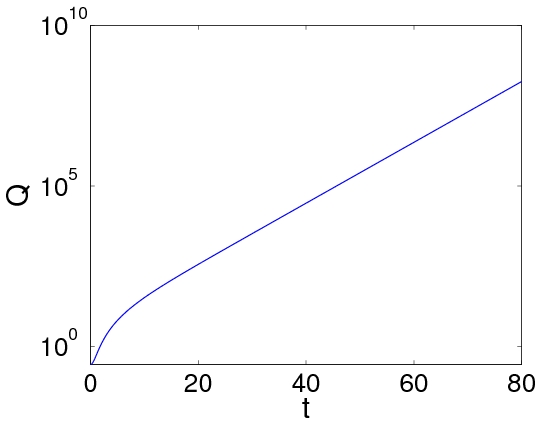} \\
\multicolumn{2}{c}{\includegraphics[width=4cm]{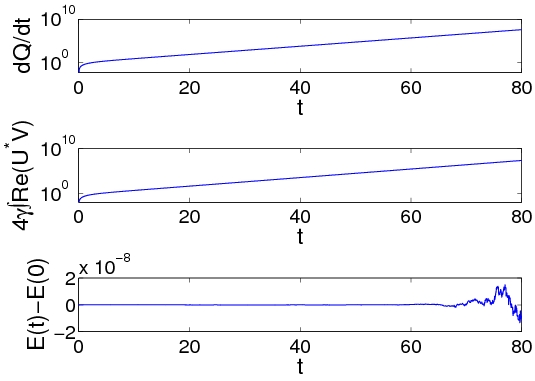}}\\
\end{tabular}
\caption{Dynamics for $\Lambda=0.8$ and $\gamma_s=1$ when the soliton with $\gamma_0=0.59$ is used as initial condition. The left panel shows both components of the spinors at different times. The right panel shows the total charge as a function of time. The lower panel compares the time derivative of the charge and the right hand side of Eq. (\ref{eq:chargederiv}) and shows the time dependence of the energy fluctuations.
One can clearly observe the (spatially independent) exponential growth of the waveform.}
\label{fig:simul3}
\end{figure}

We now turn to  the consideration of the dynamical evolution of solitons past the {\cP\cT} transition point. As indicated above, given their generic stability
for $\gamma < \gamma_{\cP \cT}$, we do not consider the latter case.
In the case of $\gamma > \gamma_{\cP \cT}$, we have firstly taken as initial
condition the soliton for $\Lambda=0.8$ and $\gamma=\gamma_0=0.59$
in the simulation with $\gamma=\gamma_s>\gPT=\sqrt{1-\Lambda^2}=0.6$. We observe that if $\gamma_s$ is close enough to $\gPT$
(i.e., for a ``shallow'' quench), the density oscillates with a frequency that decreases with $\gamma_s-\gPT$ (see Fig. \ref{fig:simul1}).
Notice that the charge of the new soliton is always higher than the charge
of the initial one (see the oscillations of
Fig. \ref{fig:simul1}) and that the maximum charge
increases with $\gamma_s$.
Interestingly, in all of these case examples we find that the
($\gamma$-independent) energy is very well conserved as is shown in
Fig~\ref{fig:simul2b}.
 When the maximum charge is above a threshold
(this occurs for $\gamma_s\gtrsim0.995$, i.e., for a deep quench), the frequency of the new soliton tends to zero and the solution starts to grow indefinitely
as shown in Fig. \ref{fig:simul3}. If a smaller value of $\gamma_0$ is
taken, the same phenomenology persists, but the indefinite growth
emerges for a smaller value of $\gamma_s$. In Fig.~\ref{fig:simul2b}
we have confirmed that both the energy conservation law and the
moment equation (\ref{eq:chargederiv})
for the power are satisfied in our dynamics.
The same is true for the case of Fig.~\ref{fig:simul3} where
the charge grows exponentially (in the case shown in the figure, for which $\gamma_s=1$, as $\sim\exp(0.088t)$; although the characteristic growth
rate depends on
$\gamma_s$). Here, the soliton does not collapse, as its shape and width
are preserved during the growth.
Again, this type of growth appears to be very different
than, say, the collapse in the Hamiltonian
NLS model~\cite{sulem}. In the latter,
the width decreases and the amplitude increases, whereas here the
entire solution grows without changing its spatial distribution.

Let us mention that the waveform with oscillating charge
is fairly generic when the quench is not sufficiently deep to
cause an exponential growth.
Remarkably, such solitary waves with
oscillating charge can be attained by performing an SU(1,1) transformation to a standing wave soliton (\ref{eq:stat}) and have
been predicted in~\cite{boubou}. This type of solution for
a standing wave of frequency $\tilde{\Lambda}$ is, in fact,
intrinsically connected to the invariance of the frequency
$2 \tilde{\Lambda}$ in the spectrum \cite{MR2892774}.
More specifically,
these bi-frequency, oscillating charge coherent structures
[which can be dubbed as SU(1,1) solitons] are of the form:
\begin{IEEEeqnarray}{rcl}\label{eq:su11}
    U(x,t) &=& \alpha_{-}\tilde{u}(x)\exp(-\ii \tilde{\Lambda} t)-\ii\alpha_{+}\tilde{v}^*(x)\exp(\ii \tilde{\Lambda} t)\,, \\
    V(x,t) &=& \alpha_{-}\tilde{v}(x)\exp(-\ii \tilde{\Lambda} t)-\ii\alpha_{+}\tilde{u}^*(x)\exp(\ii \tilde{\Lambda} t)\,,
\end{IEEEeqnarray}
with
\begin{equation}
    \alpha_{\pm}\in\mathbb{C},
    \qquad
    |\alpha_{-}|^2-|\alpha_{+}|^2=1 .
\end{equation}

In this case, $\{\tilde{u}(x),\tilde{v}(x)\}$ is the standing wave solution
with frequency $\tilde{\Lambda}$. Consequently, the charge oscillates with a frequency $2 \tilde{\Lambda}$ as long as $\gamma\neq0$. There is an SU(1,1) family of solutions for each value of $\gamma$ and $\tilde{\Lambda}$ which fulfills the same equations that the standing wave solutions (\ref{eq:stat}) satisfy.
As a result, when $\gamma_s\neq\gamma_0$, an SU(1,1) solution with $\gamma=\gamma_s$ is apparently dynamically manifested.
Since these periodic SU(1,1) solutions and the standing wave solutions
only exist for $\gamma < m \equiv 1$, it is natural to expect that
there are no nontrivial fixed points for the dynamics for
$\gamma > 1$, hence giving rise to the observed growth dynamics.

\begin{figure}
\begin{tabular}{cc}
{\footnotesize $\gamma=0$} & {\footnotesize $\gamma=0.2$} \\
\includegraphics[width=4cm]{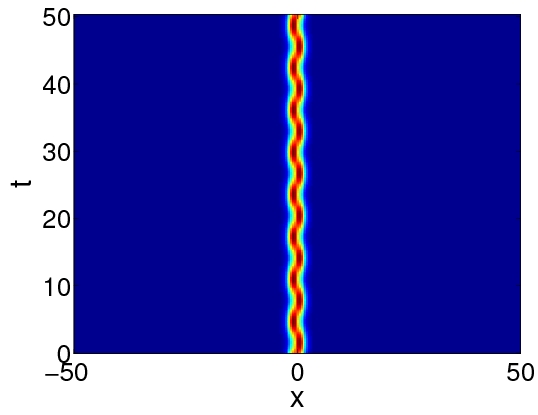} &
\includegraphics[width=4cm]{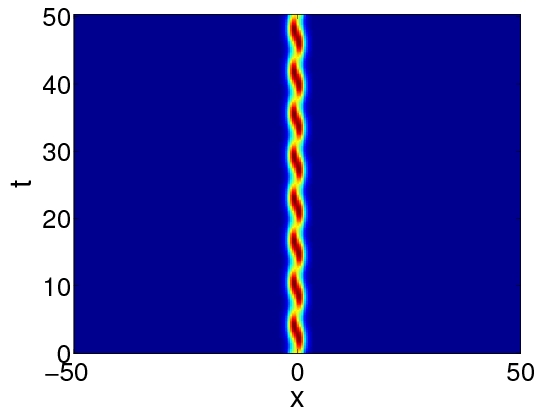} \\
{\footnotesize $\gamma=0.4$} & {\footnotesize $\gamma=0.8$} \\
\includegraphics[width=4cm]{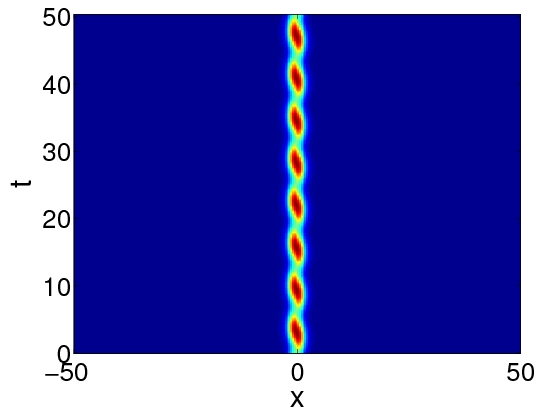} &
\includegraphics[width=4cm]{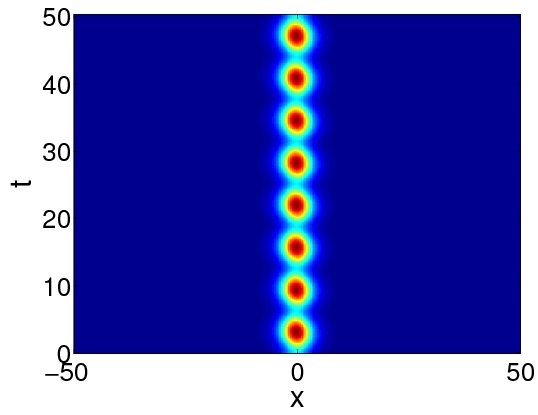} \\
\end{tabular}
\begin{tabular}{cc}
{\footnotesize $\gamma=0$} & {\footnotesize $\gamma=0.2$} \\
\includegraphics[width=4cm]{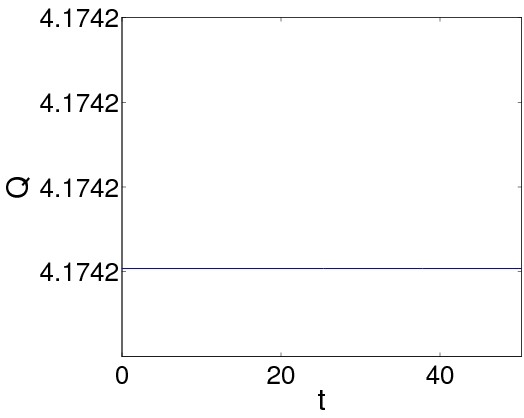} &
\includegraphics[width=4cm]{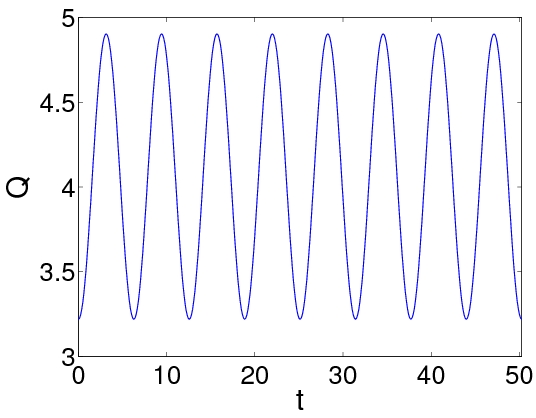} \\
{\footnotesize $\gamma=0.4$} & {\footnotesize $\gamma=0.8$} \\
\includegraphics[width=4cm]{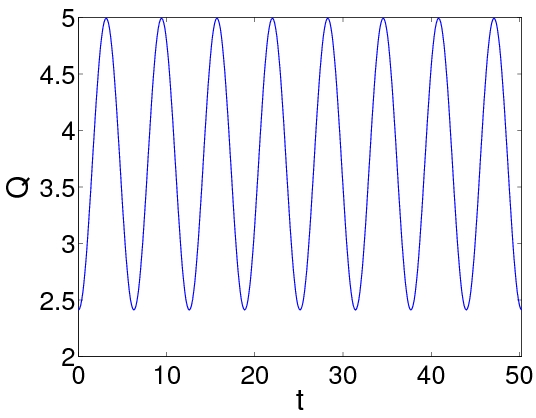} &
\includegraphics[width=4cm]{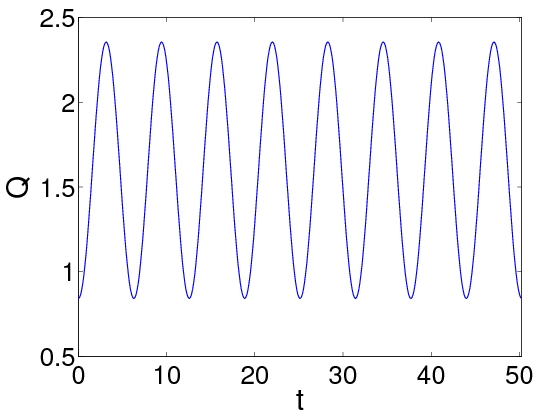} \\
\end{tabular}
\caption{Dynamics for SU(1,1) solitons with $\tilde{\Lambda}=0.5$, $\alpha_{-}=1.0500$ and $\alpha_{+}=0.3202$. Each panel shows the time evolution of the soliton density. The bottom four panels show the temporal evolution of
the charge for each of the examples in the top panels.}
\label{fig:su11}
\end{figure}

We have confirmed that the dynamics observed, e.g., in
Fig.~\ref{fig:simul1}
corresponds to SU(1,1) solutions. For instance, for the soliton with
$\gamma_0=0.59$, $\Lambda=0.8$, when initializing it
for $\gamma_s=0.9$, it spontaneously gives rise to an oscillatory state
of the above form of Eq.~(\ref{eq:su11}) with
$\tilde\Lambda=0.422$, $\alpha_{-}=1.0847$ and $\alpha_{+}=0.4201$.
On the other hand, using a numerically exact (up to a prescribed tolerance)
solution of our fixed point iteration scheme with a given frequency
$\tilde{\Lambda}$ (for a desired $\gamma$), we can select values
of  $\alpha_{-}$ and $\alpha_{+}$ and the exact form of Eq.~(\ref{eq:su11})
in order to construct, at will, such bi-frequency SU(1,1) solutions.
An example of this form is shown in
the panels of Fig.~\ref{fig:su11}  (even for $\gamma=0$)
for $\tilde{\Lambda}=0.5$, $\alpha_{-}=1.0500$ and $\alpha_{+}=0.3202$.

\subsection{Massless NLD equation with $k=1$}

Finally, we now turn to the analysis of the solutions in the special case
of $m=0$.
This case has the peculiarity that it features exact solutions
available in analytical form even for $\gamma\neq0$. Due to the functional
form of those solutions, we had to use a Chebyshev collocation scheme with
inhomogenous Dirichlet boundary conditions (to numerically identify
their stability). As we will see below
directly from their functional form,
these solutions are not \cP\cT-symmetric, hence the nonlinearity
becomes responsible for the breaking of \cP\cT-symmetry.
This, in turn, implies that the linear stability eigenvalues do not come
in quartets [but rather only in conjugate pairs i.e., in pairs
with opposite Re$(\omega)$]. Furthermore, although the solutions are stable
for $\gamma=0$, they become immediately unstable, as soon as
$\gamma\neq0$.

There are two main exact solutions for a given value of $\gamma$:

\begin{IEEEeqnarray}{rcl}\label{sol1}
    u_1(x)&=&\frac{1}{2\sqrt{-g\Lambda}}\left[(-\Lambda+\rho\tanh(\rho x))-\ii\gamma\right], \IEEEyesnumber \IEEEyessubnumber \\
    v_1(x)&=&\frac{1}{2\sqrt{-g\Lambda}}\left[(-\Lambda-\rho\tanh(\rho x))-\ii\gamma\right],  \IEEEyessubnumber
\end{IEEEeqnarray}
and
\begin{IEEEeqnarray}{rcl}\label{sol2}
    u_2(x)&=&\frac{1}{2\sqrt{-g\Lambda}}\left[(\rho+\Lambda\tanh(\rho x))+\ii\gamma\tanh(\rho x)\right], \IEEEyesnumber \IEEEyessubnumber \\
    v_2(x)&=&\frac{1}{2\sqrt{-g\Lambda}}\left[(-\rho+\Lambda\tanh(\rho x))+\ii\gamma\tanh(\rho x)\right], \IEEEyessubnumber
\end{IEEEeqnarray}
with $\rho=\sqrt{\gamma^2+\Lambda^2}$. This pair of solutions can be transformed to another pair by virtue of the transformation $\{u,v\}\rightarrow\{v^*,-u^*\}$, as Eq.~(\ref{eq:dyn}) is invariant under transformations $\{U,V\}\rightarrow\{V^*,-U^*\}$ when $m=0$. This transformation introduces the change $\omega\rightarrow\omega^*$ into the stability eigenfrequencies spectrum.
{It is worth noticing that both solutions (\ref{sol1}) and (\ref{sol2})
correspond to the same density, i.e.}

\begin{equation}
    |u_1(x)|^2+|v_1(x)|^2=|u_2(x)|^2+|v_2(x)|^2
\end{equation}

\begin{figure}
\begin{center}
\includegraphics[width=4cm]{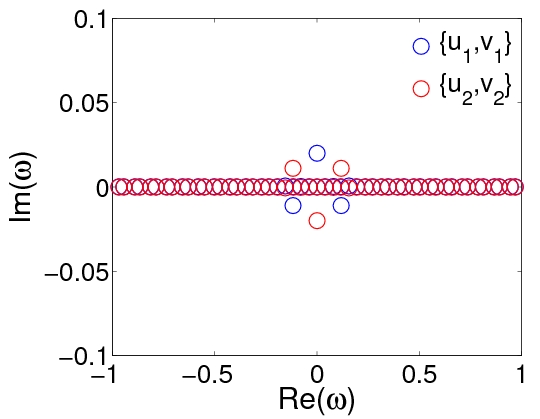}
\end{center}
\caption{Spectral plane of the
exact analytical solutions of Eqs.~(\ref{sol1})-(\ref{sol2})
for $m=0$, $\Lambda=-0.9$ and $\gamma=0.01$. The spectra
of the two solutions are conjugate, but both reflect their
non-{\cP \cT}-symmetric nature in the absence of eigenfrequency
quartets within the spectral plane.}
\label{fig:stabm0}
\end{figure}

Figure \ref{fig:stabm0} shows the spectral plane for both solutions of the massless equation for $\gamma=0.01$ and $\Lambda=-0.9$. It can be observed that the stability eigenfrequencies for the first solution are the complex conjugates
of the ones of the second solution. Nevertheless, in neither case
does the spectrum present a four-fold symmetry, given the nonlinearity-induced
breaking of {\cP \cT}-symmetry.

\begin{figure}
\begin{tabular}{cc}
\includegraphics[width=4cm]{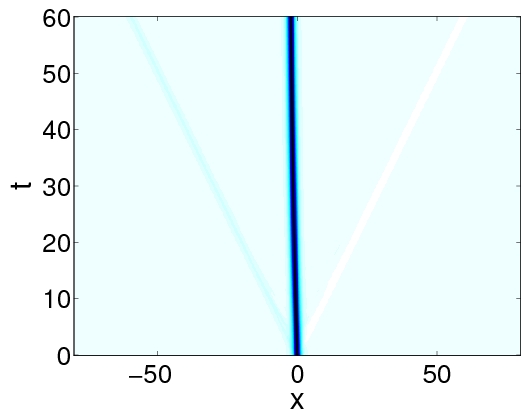} &
\includegraphics[width=4cm]{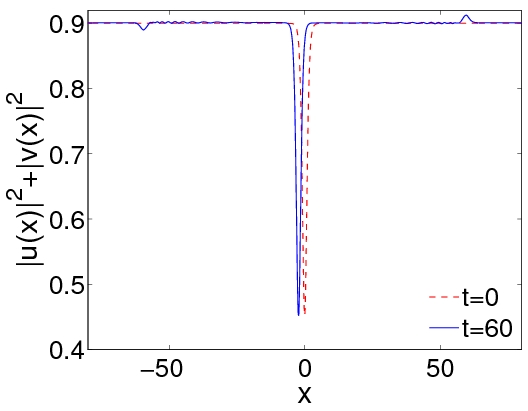} \\
\end{tabular}
\caption{Dynamics of the soliton (\ref{sol1}) for $m=0$, $\Lambda=-0.9$ and $\gamma=0.01$. Left panel shows the density plot (darker color means deeper amplitude) whereas right panel compares the initial density with that of an intermediate time.}
\label{fig:simulm0}
\end{figure}

{We have performed simulations in order to observe the effect of unstable modes for both $\{u_1,v_1\}$ and $\{u_2,v_2\}$ solutions. Those simulations have been performed using a Chebyshev collocation method with Neumann boundary conditions using a 2nd-3rd order Runge-Kutta integrator supplemented by a trapezoidal rule and the backward differentiation formula of order 2 (TR-BDF2 algorithm) \cite{TR-BDF2}. We have used as initial condition the standing wave perturbed along its unstable eigenmode corresponding to the imaginary eigenfrequency direction. As a result of the simulation (see Fig. \ref{fig:simulm0}), we can observe that the soliton dip starts to slowly move along the system with constant speed and preserving its shape whereas a counter-propagating pair of small amplitude waves is emitted. This behaviour is found for both $\{u_1,v_1\}$ and $\{u_2,v_2\}$ solutions. If the standing waves were perturbed following the direction of the eigenmode corresponding to the complex eigenvalue pair, the density
dip would remain at rest and radiation in form of a staggered perturbation is emitted; in fact, the eigenmode is spurious because its spatial
profile possesses a zigzag tail which is not compatible with solutions
in the continuum limit, hence we do not consider it further here.}

\section{Conclusions and Future Challenges}

In the present work, we introduced a {\cP \cT}-symmetric
extension of the Gross-Neveu (or Soler) model as a prototypical
example of a {\cP \cT}-symmetric nonlinear Dirac equation
(PT-NLDE).
We have found that the model possesses a number of
remarkable and previously unexplored (some even in the
standard Hamiltonian case) characteristics.
The {\cP \cT}-symmetric variant used here involved
the Dirac matrix $\gamma_5$
(whose role
in the case of spinors with two components
was played by the Pauli matrix $\sigma_1$)
that, as we argued,
introduced an {\it unusual} form of a {\cP \cT}-symmetric
setting where for our two-component model, it was not transparent
that one field bears gain and the other loss, but both, in principle,
carry both gain and loss. Perhaps this special nature of the
problem is responsible for the remarkable feature that the
Hamiltonian form of the energy (which is $\gamma$-independent)
is still conserved in the presence of the gain-loss parameter
$\gamma$ (while the power is not). On the other hand, the
nonlinear solutions identified here presented an unusual
{\cP \cT} phase transition (for the case of unit mass)
whereby they degenerated into the linear limit of
the problem. The solutions (which presented an implicit
mono-parametric dependence on $\gamma^2 + \Lambda^2$)
were found to be stable everywhere within the regime
of exact {\cP \cT}-symmetry hence we attempted to consider dynamics
beyond this interval. There, we found the prototypical formation
of time-periodic solutions whose bi-frequency character was
attributed to the SU(1,1) symmetry of the equation. Importantly
such periodic solutions exist even in the Hamiltonian limit
of the model. Beyond a critical value of $\gamma$, the
solutions were found to feature an unusual (spatially independent)
growth. Finally, exact analytical waveforms could only be identified
in the special limit of the massless problem with $m=0$. These solutions
had an unexpected characteristic as well in that the nonlinearity
enabled their breaking of the {\cP \cT}-symmetry for all values
of $(\Lambda,\gamma)$ for which they were found to exist.

Clearly, this is an unusual and exceptionally rich class of models
at the interface of the emerging theme of {\cP \cT}-symmetric
media and the mathematically highly nontrivial realm of the
nonlinear Dirac equations. Understanding in more depth any one
of the above features [the energy independence on $\gamma$,
the exponential, spatially independent growth, the spectrum of
the massless solutions or that of the SU(1,1) solutions] would
already enable significant advances in this nascent field
of {\cP \cT}-NLDE models. Generalizations including variants of
the model [such as the integrable Thirring model, or the
physically motivated self-interacting model (i.e., only
featuring respectively nonlinear terms of the form $|U|^2 U$
and $|V|^2 V$ in Eq.~(\ref{eq:dyn}))]
would be particularly worthwhile to consider,
as would extensions to other settings such as two-dimensional
ones. Such extensions are presently under consideration and relevant
results will be reported in future publications.

\section*{Acknowledgement}

The work of PGK, AS and FC at Los Alamos is partially supported
by the US Department of Energy.
P.G.K.~gratefully acknowledges the support of
NSF-DMS-1312856, BSF-2010239, as well as from
the US-AFOSR under grant FA9550-12-1-0332,
and the ERC under FP7, Marie Curie Actions, People,
International Research Staff Exchange Scheme (IRSES-605096).
The research of AC
was carried out at the Institute for Information Transmission Problems
of the Russian Academy of Sciences at the expense of the Russian Foundation
for Sciences (project 14-50-00150).
AK wishes to thank Indian National Science Academy (INSA) for the
award of INSA Senior Scientist Position.

\begin{IEEEbiography}[{\includegraphics[width=1in,height=1.25in,clip,keepaspectratio]{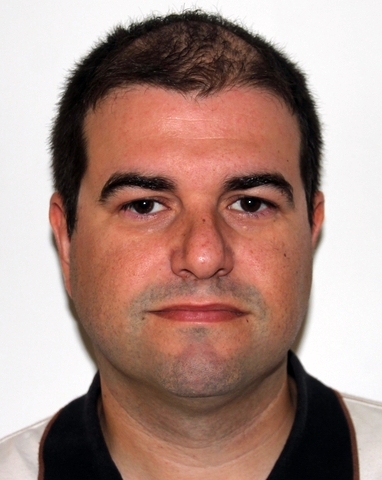}}]{Jes\'us Cuevas--Maraver}
received the B.Sc. in Physics from the University of Sevilla, Seville, Spain, in 1999. Subsequently, he become member of the Nonlinear Physics Group at the same university and in 2000, he joined the Applied Physics Department, receiving the PhD in Physics in 2003 for his work in intrinsic localized modes in nonlinear lattices. In May 2009 he become {\em Profesor Titular} (Associate Professor) at the same department and from 2014 is member of the Mathematics Institute of the University of Seville (IMUS). His main research interest comprises the study of localized waves in nonlinear media, ranging from photonic systems to biomolecules or Bose-Einstein condensates.
\end{IEEEbiography}

\begin{IEEEbiography}[{\includegraphics[width=1in,height=1.25in,clip,keepaspectratio]{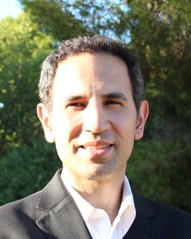}}]{Panayotis G. Kevrekidis}
earned his B.Sc. from the University of Athens (1996, Physics). He joined the Physics Department of Rutgers earning his M.S. (1998), M.Phil. and Ph.D. (2000) in Physics. He then assumed a post-doctoral post split between the Program in Applied and Computational Mathematics of Princeton University  and the Theoretical Division and the Center for NonLinear Studies of the Los Alamos National Laboratory in 2000-2001. From 09/2001, he joined the Department of Mathematics and Statistics of the University of Massachusetts (UMass), Amherst, as an Assistant Professor. He became an Associate Professor at UMass in 2005 and a Full Professor in 2010.
\end{IEEEbiography}

\begin{IEEEbiography}[{\includegraphics[width=1in,height=1.25in,clip,keepaspectratio]{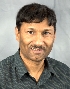}}]{Avadh Saxena}
is Group Leader of the Condensed Matter and Complex Systems
group at the Theoretical Division at Los Alamos National Laboratory, New Mexico,
USA.  He is also an affiliate of the Center for Nonlinear Studies at Los Alamos. He
received his Ph. D. in Physics from Temple University, Philadelphia in 1986 and
after a postdoc at the Pennsylvania State University joined Los Alamos in 1990.
His expertise is in condensed matter physics, phase transitions in functional
materials a variety of nonlinear phenomena.
\end{IEEEbiography}

\begin{IEEEbiography}[{\includegraphics[width=1in,height=1.25in,clip,keepaspectratio]{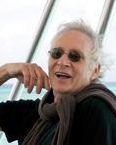}}]{Fred Cooper}
completed his studies in Physics from City College New York in 1964, and received his Ph. D. in Physics from Harvard University in 1968. He was an Instructor at Cornell University from 1968-1970, an Assistant Professor of Physics at Belfer Graduate School of Science from 1970-1975
abad a Staff Member at Los Alamos National Labs (LANL)  in the Theory division from 1975-2012, and was Group Leader of the Elementary Particle Physics group T-8 from 1995-2001.
He was Program Director for Theoretical Physics at the National Science Foundation from 2002-2009
He currently is an external Professor at the Santa Fe Institute and a guest scientist at LANL.
His expertise is in Elementary Particle Physics, the theory of Bose Einstein Condensates, Nonlinear dynamics, and Pattern Formation in Chemical reactions. He is a member of the APS.
\end{IEEEbiography}

\begin{IEEEbiography}[{\includegraphics[width=1in,height=1.25in,clip,keepaspectratio]{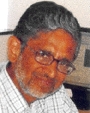}}]{Avinash Khare}
got his Masters in Physics from Holkar college Indore, India
and his Ph.D. in Theoretical High Energy Physics from Saha Institute of
Nuclear Physics, Kolkata, India in 1971. After a postdoctoral position at
University of Tokyo, he moved to Institute of Physics, Bhubaneswar where
he retired as Senior Professor in 2010. Since then he worked as Raja
Ramanna Fellow at Indian Institute of Science Education and Research, Pune
and in June 2015 he has moved to Physics Department, Savitribai Phule Pune
University, Pune, India as INSA Senior Scientist.
\end{IEEEbiography}

\begin{IEEEbiography}[{\includegraphics[width=1in,height=1.25in,clip,keepaspectratio]{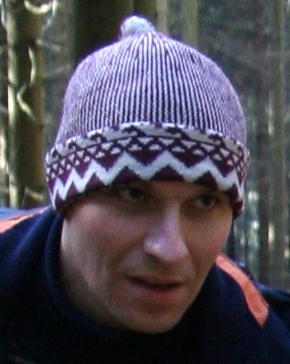}}]{Andrew Comech}
completed his M.S. studies in Theoretical Physics
at Moscow Institute for Physics and Technology in 1993,
and received his Ph.D. in Mathematics from Columbia University in 1997.
He is currently an Associate Professor at Mathematics Department at
Texas A\&M University, working on stability properties of solitary
wave solutions.
\end{IEEEbiography}

\begin{IEEEbiography}[{\includegraphics[width=1in,height=1.25in,clip,keepaspectratio]{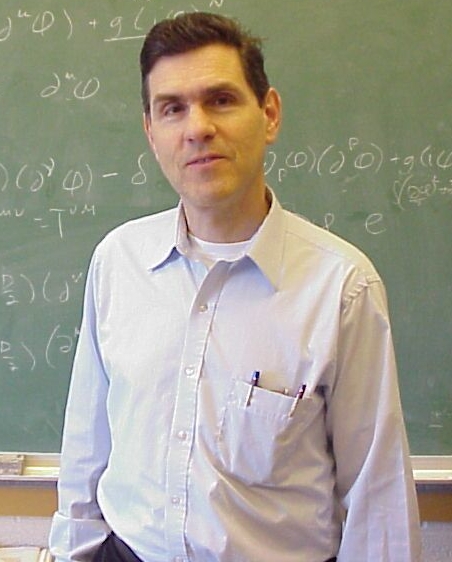}}]{Carl M. Bender}
got his undergraduate degree in physics from Cornell University and his PhD in theoretical physics from Harvard University in 1969. After a postdoctoral position at the Institute for Advanced Study (Princeton) and Assistant and Associate Professorships at MIT, he moved to Washington University in St. Louis, where he is the Wilfred R. and Ann Lee Konneker Distinguished Professor of Physics.
\end{IEEEbiography}

\end{document}